\documentclass{article}

\usepackage{microtype}
\usepackage{graphicx}
\usepackage{subfigure}
\usepackage{booktabs} 

\usepackage{hyperref}

\usepackage[accepted]{icml2023}

\usepackage{amsmath}
\usepackage{amssymb}
\usepackage{mathtools}
\usepackage{amsthm}

\usepackage[capitalize,noabbrev]{cleveref}

\theoremstyle{plain}

\theoremstyle{definition}

\theoremstyle{remark}

\usepackage[textsize=tiny]{todonotes}

\icmltitlerunning{Differentiating Through Linear Solvers}

\begin{document}

\twocolumn[
\icmltitle{Differentiating Through Linear Solvers}



\icmlsetsymbol{equal}{*}

\begin{icmlauthorlist}
\icmlauthor{Jan H\"uckelheim}{equal,yyy}
\icmlauthor{Paul Hovland}{equal,yyy}
\end{icmlauthorlist}

\icmlaffiliation{yyy}{Argonne National Laboratory, Lemont, IL, USA}

\icmlcorrespondingauthor{Jan H\"uckelheim}{jhueckelheim@anl.gov}

\icmlkeywords{Automatic Differentiation, Linear Solvers}

\vskip 0.3in
]



\printAffiliationsAndNotice{\icmlEqualContribution} 

\begin{abstract}
Computer programs containing calls to linear solvers are a known challenge for automatic differentiation. Previous publications advise against differentiating through the low-level solver implementation, and instead advocate for high-level approaches that express the derivative in terms of a modified linear system that can be solved with a separate solver call.
Despite this ubiquitous advice, we are not aware of prior work comparing the accuracy of both approaches. With this article we thus empirically study a simple question: What happens if we ignore common wisdom, and differentiate through linear solvers?
\end{abstract}

\section{Introduction}
\label{sec:intro}

One of the merits of automatic, or algorithmic, differentiation (AD or autodiff) is that it is applied directly to algorithms, applying rule-based differentiation to the low-level functions intrinsic to a particular programming language and combining those derivatives according to the chain rule of differential calculus~\cite{griewank2008evaluating}.  
Nonetheless, when applying AD to programs including iterative solvers, conventional wisdom advises treating the solver as an intrinsic function, rather than ``naively'' differentiating through the low-level implementation details.  This approach is particularly common in the case of fixed point iterations, where there is a wealth of established theory~\cite{christianson1994reverse,griewank1993derivative} and many AD tools support high-level differentiation of fixed-point iterative solvers that have labeled as such.  

The situation for linear solvers is more complicated.  While stationary iterative methods adhere to the theory of fixed point iterations, Krylov methods do not, and the theory in this regime is much less developed. Gratton et al.~\yrcite{gratton2014differentiating} and Christianson~\yrcite{christianson2018differentiating} have advanced the theory for conjugate gradient solvers (see Section~\ref{sec:background} for more details). However, it is unclear how much of this theory carries over to solvers for nonsymmetric systems, such as GMRES and BiCGStab.  Furthermore, even in cases where theory argues for high-level differentiation, practical considerations such as difficulty in clearly delineating the solver interface or problems finding an appropriate preconditioner for the transposed system might argue in favor of low-level differentiation.

This paper examines empirically the tradeoffs between low-level differentiation and high-level differentiation of several linear solvers, as implemented in SPARSKIT~\cite{saad1994sparskit}. The remainder of this paper is organized as follows. The next section provides additional background on low-level and high-level differentiation of linear solvers. Section~\ref{sec:exp-design} discusses our experimental design, followed by our experimental results in Section~\ref{sec:exp-results}.  We conclude with a summary of our findings and our plans for future work.

\section{Background}
\label{sec:background}

Automatic differentiation has two main ``modes,'' forward (or tangent) and reverse (or adjoint) mode.  In the so-called forward mode, the chain rule is applied starting with the independent variables and propagating derivatives forward to the dependent variables. In the so-called reverse mode, the chain rule is applied starting with the dependent variables and propagating backward to the independent variables. For this paper, we will focus primarily on the forward mode.

Conjugate gradient is the best-studied Krylov solver with respect to low-level differentiation.  Gratton et al.~\yrcite{gratton2014differentiating} examined the consequences of differentiating through a conjugate gradient (CG) solver, under the assumption that all of the iterates $x_k$ are well-defined, that is, CG does not reach the exact solution $A^{-1}b$ before iteration $k$.  Christianson~\yrcite{christianson2018differentiating} also considered low-level differentiation of CG, but in the case where the exact solution is in fact found, proving that in exact arithmetic and with appropriate modifications, both the solution to the linear system and its derivatives can be computed in no more than $N$ iterations, for $A \in \mathcal{R}^{N\times N}$.  Low-level differentiation of Krylov solvers for nonsymmetric systems is at present poorly understood at a theoretical level.

The alternative to low-level differentiation is high-level differentiation, treating the linear solve as an elementary function and applying the rules of matrix calculus to compute the derivatives directly.  The derivation of such rules is straightforward for the forward mode.  For independent variable $u$ and since $Ax = b$, we have
\begin{equation*}
    \frac{\partial A}{\partial u}x + A\frac{\partial x}{\partial u} = \frac{\partial b}{\partial u}.
\end{equation*}
Consequently, one can compute $\frac{\partial x}{\partial u}$ by solving the linear system $Ay = \hat{b}$, where $\frac{\partial x}{\partial u} = y$ and
\begin{equation*}
    \hat{b} = \frac{\partial b}{\partial u} - \frac{\partial A}{\partial u}x.
\end{equation*}
Because the matrix $A$ is used in both linear solves, we can reuse the preconditioner used to solve for $x$.  The situation is more complicated for the reverse mode and nonsymmetric matrices, where the linear solve used to compute derivatives involves the matrix transpose, $A^T$.

\section{Experimental Setup}
\label{sec:exp-design}

We implemented low-level differentiation by applying the Tapenade AD tool~\cite{hascoet2013tapenade} to the SPARSKIT implementation of GMRES, TFQMR and BiCGStab. We modified the SPARSKIT implementation to replace Fortran language features that caused compilation errors after differentiation with Tapenade, such as computed and assigned goto statements, and created a head routine that implements the choice of linear solver with a branch instead of passing the name of an external subroutine as an argument. We also replaced the BLAS implementation in SPARSKIT with a more recent version~\cite{blackford2002updated}. We then implemented high-level differentiation of these solvers at the matrix calculus level, following the derivation in Section~\ref{sec:background}.

We evaluated both approaches using 65 matrices from the SuiteSparse collection~\cite{10.1145/2049662.2049663} (see~\ref{sec:matrixlist} for the list), comprising all of the nonsymmetric matrices from the Bai collection with up to $1,000$ rows and columns.  To facilitate our assessment of the effectiveness of both strategies, we manufactured the righthand sides $b$ and $\frac{\partial b}{\partial u}$ from known $x$ and $\frac{\partial x}{\partial u}$, each with elements drawn from the uniform distribution $[0,1)$.  We monitored the progress after $k$ iterations, for $k \in [1,50]$, computing the L2 norm of the difference between the computed $x$ (or $\frac{\partial x}{\partial u}$) and the reference value. 

\section{Experimental Results}
\label{sec:exp-results}

\begin{figure*}[btp]
    \centering
\includegraphics[width=0.95\textwidth]{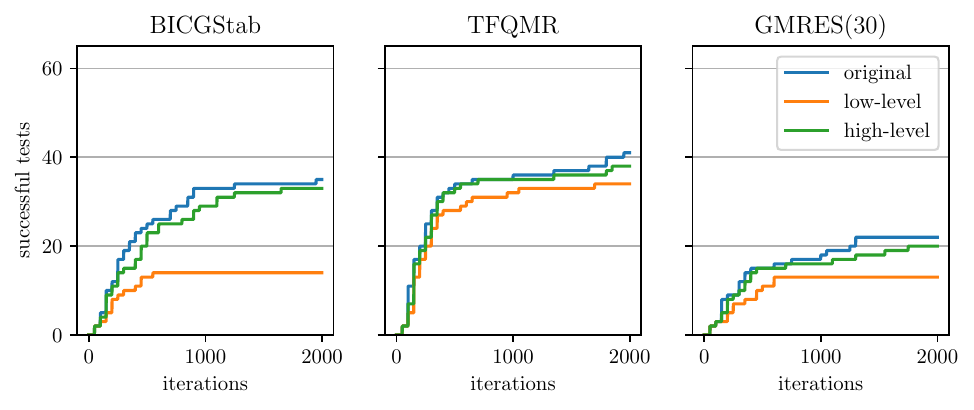}
    \caption{Data profile showing the relative performance of the low-level and high-level differentiation strategies.  The curve for the original linear solver shows the number of problems solved (out of 65) such that $\left\|x - x_{\mbox{\scriptsize ref}}\right\|_2 < 10^{-2}$.  The curves for the differentiated linear solvers show the number of problems solved such that $\left\|\frac{\partial x}{\partial u} - \left(\frac{\partial x}{\partial u}\right)_{\mbox{\scriptsize ref}}\right\|_2 < 10^{-2}$.}
    \label{fig:bai2}
\end{figure*}

\begin{figure*}
    \centering
\includegraphics[width=0.95\textwidth]{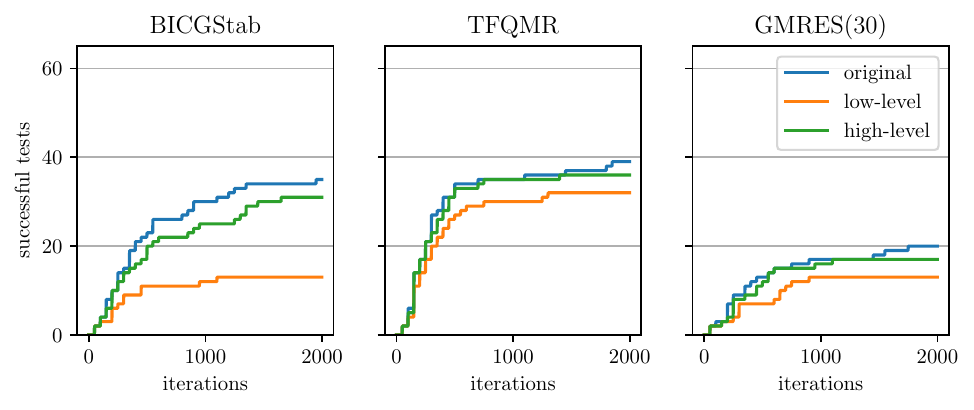}
    \caption{Data profile showing the relative performance of the low-level and high-level differentiation strategies.  The curve for the original linear solver shows the number of problems solved (out of 65) such that $\left\|x - x_{\mbox{\scriptsize ref}}\right\|_2 < 10^{-4}$.  The curves for the differentiated linear solvers show the number of problems solved such that $\left\|\frac{\partial x}{\partial u} - \left(\frac{\partial x}{\partial u}\right)_{\mbox{\scriptsize ref}}\right\|_2 < 10^{-4}$.}
    \label{fig:bai4}
\end{figure*}

Figures~\ref{fig:bai2} and~\ref{fig:bai4} show the effectiveness of the two differentiation strategies in comparison with the original, undifferentiated linear solver.  For the original solver, the curve shows the number of problems solved (out of 65) such that the L2 norm of the difference between the computed $x$ and the reference value for $x$ is less than some threshold ($10^{-2}$ or $10^{-4}$).  For the two differentiation strategies, the curves show the number of problems solved such that the L2 norm of the difference between the computed $\frac{\partial x}{\partial u}$ an the reference value is less than the same threshold.  A higher curve indicates that more problems are solved within a given number of iterations and a curve further to the left indicates that a given number of problems are solved in fewer iterations.

A few patterns emerge from our results.  First, none of the linear solvers is able to solve more than two thirds of the problems to within $10^{-4}$ of the reference solution in 2000 iterations, and restarted GMRES significantly underperforms TFQMR and BiCGStab.  Second, in general high-level differentiation performs nearly as well as the original solver, but there are typically a few problems that require more iterations to achieve similar levels of accuracy. Third, the effectiveness of low-level differentiation is highly solver-dependent.  For TFQMR and restarted GMRES, the low-level differentiation strategy is nearly as effective as high-level differentiation.  In contrast, for BiCGStab, there is a sizable gap in performance between high-level and low-level differentiation, with the latter typically only able to solve half as many problems within a similar number of iterations.

\begin{figure*}
    \centering
    \includegraphics{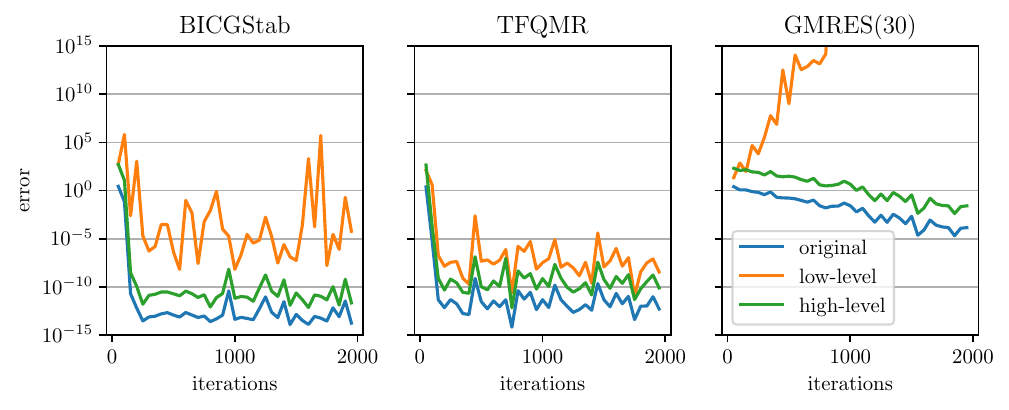}
    \caption{Convergence for the original solver and both differentiation strategies when applied to the BFWA62 matrix. The curves show the error in the system solution as a function of the iterations performed, where a lower value is better. The BICGStab and TFQMR solvers rapidly reach a value close to machine precision and remain relatively stable at that lavel. GMRES converges more slowly. High-level differentiation broadly follows this trend for all solvers. Low-level differentiation performs more erratically, and consistently underperforms high-level differentiation. Nevertheless, low-level differentiation appears to work reasonably well for TFQMR.}
    \label{fig:bfwa62}
\end{figure*}

\begin{figure*}
    \centering
    \includegraphics{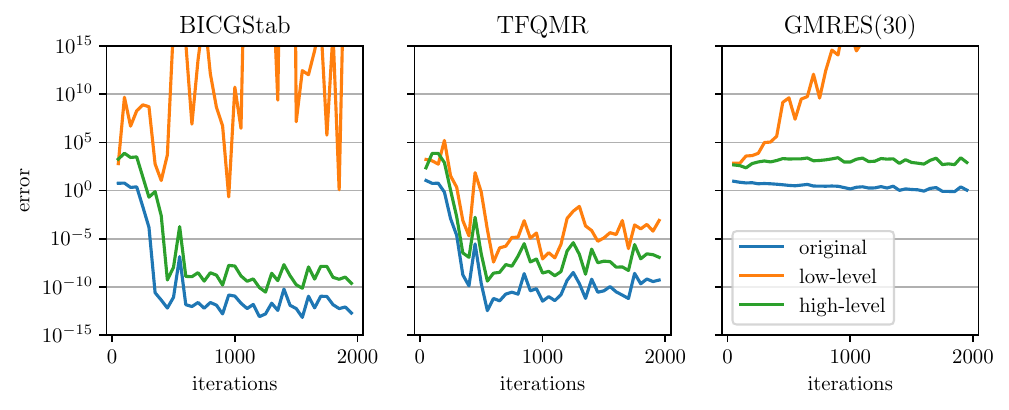}
    \caption{Convergence for the original solver and both differentiation strategies when applied to the BFWA398 matrix. Similar to the results in Figure~\ref{fig:bfwa62}, BICGStab and TFQMR rapidly converge and outperform GMRES. High-level differentiation performs worse than the original solver, but far better than low-level differentiation. Once again, TFQMR appears to be better suited for low-level differentiation and leads to reasonable results for both differentiation approaches.}
    \label{fig:bfwa398}
\end{figure*}

Figures~\ref{fig:bfwa62} and~\ref{fig:bfwa398} show detailed results for two matrices, namely BFWA398 and BFWA62 from the collection. These matrices emerge from computational electromagnetics problems, and have reasonable condition numbers of $2.993111e+03$ and $5.530615e+02$, respectively. The original linear solvers therefore converge well, as do the high-level differentiated solvers. Low-level differentiation through the solvers leads to very erratic behavior for BiCGStab, and divergence for GMRES in both cases. For TFQMR both differentiation strategies appear to work well.

\section{Conclusions}

We demonstrate with this article that the common advice is justified, and that high-level differentiation is indeed usually preferable to low-level differentiation. However, our work also shows a more nuanced picture: some linear solvers appear to be better suited for low-level differentiation than others, and in situations where high-level differentiation is not desirable -- for example due to the manual development effort that is often involved -- a careful solver choice may lead to useful gradients even with low-level approaches.

In future work, we plan to investigate reverse mode differentiation through solvers, extend the collection of solvers and preconditioners, and compare empirical results with the theory presented in previous work~\cite{christianson2018differentiating}. We are also interested in quantifying the influence of roundoff errors on our results.

\section{Acknowledgements}
This work was partly supported by the Applied Mathematics activity within the U.S. Department of Energy, Office of Science, Advanced Scientific Computing Research Program, under contract number DE-AC02-06CH11357.

\bibliography{main}
\bibliographystyle{icml2023}

\newpage
\appendix
\onecolumn
\section{List of SuiteSparse Matrices}
\label{sec:matrixlist}

The following are the SuiteSparse~\cite{10.1145/2049662.2049663} matrices used in our experiments.

\begin{table}[!h]
\centering
\scalebox{0.7}{
\begin{tabular}{llllllll}
\textbf{Id} & \textbf{Name} & \textbf{Group} & \textbf{Rows} & \textbf{Cols} & \textbf{Nonzeros} & \textbf{Kind}                                   & \textbf{Date} \\\hline
292         & bfwa398       & Bai            & 398           & 398           & 3,678             & Electromagnetics Problem                        & 1994          \\
293         & bfwa62        & Bai            & 62            & 62            & 450               & Electromagnetics Problem                        & 1994          \\
294         & bfwa782       & Bai            & 782           & 782           & 7,514             & Electromagnetics Problem                        & 1994          \\
298         & bwm200        & Bai            & 200           & 200           & 796               & Chemical Process Simulation Problem             & 1992          \\
299         & bwm2000       & Bai            & 2,000         & 2,000         & 7,996             & Chemical Process Simulation Problem             & 1992          \\
300         & cdde1         & Bai            & 961           & 961           & 4,681             & Computational Fluid Dynamics Problem Sequence   & 1992          \\
301         & cdde2         & Bai            & 961           & 961           & 4,681             & Subsequent Computational Fluid Dynamics Problem & 1992          \\
302         & cdde3         & Bai            & 961           & 961           & 4,681             & Subsequent Computational Fluid Dynamics Problem & 1992          \\
303         & cdde4         & Bai            & 961           & 961           & 4,681             & Subsequent Computational Fluid Dynamics Problem & 1992          \\
304         & cdde5         & Bai            & 961           & 961           & 4,681             & Subsequent Computational Fluid Dynamics Problem & 1992          \\
305         & cdde6         & Bai            & 961           & 961           & 4,681             & Subsequent Computational Fluid Dynamics Problem & 1992          \\
306         & ck104         & Bai            & 104           & 104           & 992               & 2D/3D Problem                                   & 1986          \\
307         & ck400         & Bai            & 400           & 400           & 2,860             & 2D/3D Problem                                   & 1986          \\
308         & ck656         & Bai            & 656           & 656           & 3,884             & 2D/3D Problem                                   & 1986          \\
309         & dw1024        & Bai            & 2,048         & 2,048         & 10,114            & Electromagnetics Problem                        & 1993          \\
310         & dw256A        & Bai            & 512           & 512           & 2,480             & Electromagnetics Problem                        & 1993          \\
311         & dw256B        & Bai            & 512           & 512           & 2,500             & Electromagnetics Problem                        & 1993          \\
312         & dw4096        & Bai            & 8,192         & 8,192         & 41,746            & Electromagnetics Problem                        & 1993          \\
313         & lop163        & Bai            & 163           & 163           & 935               & Statistical/Mathematical Problem                & 1978          \\
314         & mhda416       & Bai            & 416           & 416           & 8,562             & Electromagnetics Problem                        & 1994          \\
316         & odepa400      & Bai            & 400           & 400           & 1,201             & 2D/3D Problem                                   & 1978          \\
318         & olm100        & Bai            & 100           & 100           & 396               & Computational Fluid Dynamics Problem            & 1994          \\
319         & olm1000       & Bai            & 1,000         & 1,000         & 3,996             & Computational Fluid Dynamics Problem            & 1994          \\
320         & olm2000       & Bai            & 2,000         & 2,000         & 7,996             & Computational Fluid Dynamics Problem            & 1994          \\
321         & olm500        & Bai            & 500           & 500           & 1,996             & Computational Fluid Dynamics Problem            & 1994          \\
322         & olm5000       & Bai            & 5,000         & 5,000         & 19,996            & Computational Fluid Dynamics Problem            & 1994          \\
323         & pde225        & Bai            & 225           & 225           & 1,065             & 2D/3D Problem                                   & 1982          \\
324         & pde2961       & Bai            & 2,961         & 2,961         & 14,585            & 2D/3D Problem                                   & 1982          \\
325         & pde900        & Bai            & 900           & 900           & 4,380             & 2D/3D Problem                                   & 1982          \\
328         & qh882         & Bai            & 882           & 882           & 3,354             & Power Network Problem                           & 1994          \\
329         & rbsa480       & Bai            & 480           & 480           & 17,088            & Robotics Problem                                & 1993          \\
330         & rbsb480       & Bai            & 480           & 480           & 17,088            & Robotics Problem                                & 1993          \\
331         & rdb2048       & Bai            & 2,048         & 2,048         & 12,032            & Computational Fluid Dynamics Problem            & 1994          \\
332         & rdb5000       & Bai            & 5,000         & 5,000         & 29,600            & Computational Fluid Dynamics Problem            & 1994          \\
333         & rdb968        & Bai            & 968           & 968           & 5,632             & Computational Fluid Dynamics Problem            & 1994          \\
334         & rw136         & Bai            & 136           & 136           & 479               & Statistical/Mathematical Problem                & 1978          \\
335         & rw496         & Bai            & 496           & 496           & 1,859             & Statistical/Mathematical Problem                & 1978          \\
336         & rw5151        & Bai            & 5,151         & 5,151         & 20,199            & Statistical/Mathematical Problem                & 1978          \\
337         & tub100        & Bai            & 100           & 100           & 396               & Computational Fluid Dynamics Problem            & 1994          \\
338         & tub1000       & Bai            & 1,000         & 1,000         & 3,996             & Computational Fluid Dynamics Problem            & 1994          \\
1612        & cryg10000     & Bai            & 10,000        & 10,000        & 49,699            & Materials Problem                               & 1996          \\
1613        & cryg2500      & Bai            & 2,500         & 2,500         & 12,349            & Materials Problem                               & 1996          \\
1614        & dw2048        & Bai            & 2,048         & 2,048         & 10,114            & Electromagnetics Problem                        & 1993          \\
1615        & dw8192        & Bai            & 8,192         & 8,192         & 41,746            & Electromagnetics Problem                        & 1993          \\
1616        & dwa512        & Bai            & 512           & 512           & 2,480             & Electromagnetics Problem                        & 1993          \\
1617        & dwb512        & Bai            & 512           & 512           & 2,500             & Electromagnetics Problem                        & 1993          \\
1620        & mhd1280a      & Bai            & 1,280         & 1,280         & 47,906            & Electromagnetics Problem                        & 1994          \\
1622        & mhd3200a      & Bai            & 3,200         & 3,200         & 68,026            & Electromagnetics Problem                        & 1994          \\
1624        & mhd4800a      & Bai            & 4,800         & 4,800         & 102,252           & Electromagnetics Problem                        & 1994          \\
1626        & qh1484        & Bai            & 1,484         & 1,484         & 6,110             & Power Network Problem                           & 1994          \\
1627        & qh768         & Bai            & 768           & 768           & 2,934             & Power Network Problem                           & 1994          \\
1628        & rdb1250       & Bai            & 1,250         & 1,250         & 7,300             & Computational Fluid Dynamics Problem            & 1994          \\
1629        & rdb1250l      & Bai            & 1,250         & 1,250         & 7,300             & Computational Fluid Dynamics Problem            & 1994          \\
1630        & rdb200        & Bai            & 200           & 200           & 1,120             & Computational Fluid Dynamics Problem            & 1994          \\
1631        & rdb200l       & Bai            & 200           & 200           & 1,120             & Computational Fluid Dynamics Problem            & 1994          \\
1632        & rdb2048\_noL  & Bai            & 2,048         & 2,048         & 12,032            & Computational Fluid Dynamics Problem            & 1994          \\
1633        & rdb3200l      & Bai            & 3,200         & 3,200         & 18,880            & Computational Fluid Dynamics Problem            & 1994          \\
1634        & rdb450        & Bai            & 450           & 450           & 2,580             & Computational Fluid Dynamics Problem            & 1994          \\
1635        & rdb450l       & Bai            & 450           & 450           & 2,580             & Computational Fluid Dynamics Problem            & 1994          \\
1636        & rdb800l       & Bai            & 800           & 800           & 4,640             & Computational Fluid Dynamics Problem            & 1994          \\
1637        & tols1090      & Bai            & 1,090         & 1,090         & 3,546             & Computational Fluid Dynamics Problem            & 1991          \\
1638        & tols2000      & Bai            & 2,000         & 2,000         & 5,184             & Computational Fluid Dynamics Problem            & 1991          \\
1639        & tols340       & Bai            & 340           & 340           & 2,196             & Computational Fluid Dynamics Problem            & 1991          \\
1640        & tols4000      & Bai            & 4,000         & 4,000         & 8,784             & Computational Fluid Dynamics Problem            & 1991          \\
1641        & tols90        & Bai            & 90            & 90            & 1,746             & Computational Fluid Dynamics Problem            & 1991 
\end{tabular}
}
\caption{Matrices used in our experiments.}
\end{table}


\end{document}